\documentclass[12pt,preprint]{aastex}
\usepackage{ulem}

\title{Lyman Break Galaxies at $z \sim 5$: Rest-Frame UV Spectra. III
\footnote{Based on observations obtained at the Gemini Observatory, 
which is operated by the Association of Universities for Research 
in Astronomy, Inc., under a cooperative agreement with the NSF 
on behalf of the Gemini partnership: the National Science Foundation (United
States), the Science and Technology Facilities Council (United Kingdom), the
National Research Council (Canada), CONICYT (Chile), 
the Australian Research Council (Australia), Minist\'{e}rio da Ci\^{e}ncia
e Tecnologia (Brazil) and SECYT (Argentina).}}
\author{Hiroki Kajino\altaffilmark{2}, Kouji Ohta\altaffilmark{2}, 
Ikuru Iwata\altaffilmark{3},Kiyoto Yabe\altaffilmark{2}, \\ 
Suraphong Yuma\altaffilmark{2}, Masayuki Akiyama\altaffilmark{4},
 Naoyuki Tamura\altaffilmark{5}, \\ Kentaro Aoki\altaffilmark{5}, 
and Marcin Sawicki\altaffilmark{6}}
\date{\empty}
\email{ohta@kusastro.kyoto-u.ac.jp}
\altaffiltext{2}{Department of Astronomy, Kyoto University, Kyoto
606-8502, Japan}
\altaffiltext{3}{Okayama Astrophysical Observatory, 
National Astronomical Observatory of Japan, Okayama 719-0232, Japan}
\altaffiltext{4}{Astronomical Institute, Tohoku University,
Sendai 980-77, Japan}
\altaffiltext{5}{Subaru Telescope, National Astronomical Observatory of
Japan, 650 North A'ohoku Place, Hilo, HI 96720}
\altaffiltext{6}{Department of Astronomy and Physics, St. Mary's
University,
923 Robie St., Halifax, Nova Scotia, B3H 3C3, Canada}
\begin{document}

\begin{abstract}
We present results of optical spectroscopic observations of candidates 
of Lyman Break Galaxies (LBGs) at $z \sim 5$ in the region 
including the GOODS-N and the J0053+1234 region by using GMOS-N and GMOS-S,
 respectively.
Among 25 candidates, five objects are identified to be at $z \sim 5$ 
(two of them were already identified by an earlier study) 
and one object very close to the color-selection 
window turned out to be a foreground galaxy.
With this spectroscopically identified sample and those from
previous studies, we derived the lower limits on the number density of bright 
($M_{UV}<-22.0$ mag) LBGs at $z \sim 5$.
These lower limits are comparable to or slightly smaller than 
the number densities of UV luminosity functions (UVLFs) 
that show the smaller number density among $z \sim 5$ UVLFs in literature.
However, by considering that there remain many LBG candidates 
without spectroscopic observations, the number density of bright LBGs is expected to increase by a factor of two or more.
The evidence for the deficiency of UV luminous LBGs with 
large Ly$\alpha$ equivalent widths was reinforced.
We discuss possible causes for the deficiency and prefer 
the interpretation of dust absorption.
\end{abstract}

\keywords{galaxies: evolution --- galaxies: formation --- galaxies: high-redshift}

\section{INTRODUCTION}
\label{introduction}
In order to understand formation and evolution of galaxies, it is necessary to search and study high-redshift galaxies. 
Lyman Break Galaxies (LBGs), which are selected by rest-frame UV broad band  photometry (e.g., \citealp{ste92,ste95}), 
make the largest sample of galaxies at $z \gtrsim 3$ among various populations selected through different methods, and their statistical and individual studies have been made extensively. 
For instance, based on photometric samples, 
rest-frame UV luminosity functions (UVLFs) of LBGs at $z \sim 3-6$ are
derived (e.g.,
\citealp{ste99,iwa03,iwa07,leh03,ouc04a,bec06,saw06,yos06,bou07}) 
and attempts to measure the LF have been made even at higher redshifts (e.g.,
\citealp{ric06,ric08,star07,bou08,sta08b,oes08}).
These studies are revealing the cosmic star formation history; the cosmic star formation rate density rises from $z \sim 5-6$ to $z \sim 2-3$ and turns to decline toward $z \sim 0$ (e.g., \citealp{hop06}).
However two different evolutions of UVLFs are claimed.
One is that from $z \sim 5$ to $\sim3$ the number density of
UV faint galaxies increases while that of bright galaxies remains
almost constant (\citealp{saw06,iwa07}).
The other is that while the number density of UV faint galaxies
remains constant, that of bright galaxies increase
(\citealp{yos06,bou07}).
The number density of UVLF in bright part is key to understand the
galaxy evolution at these redshifts.

Follow-up optical spectroscopic surveys have also been made 
(e.g., \citealp{ste96a, ste96b,ste99}).
\citet{sha03} studied $\sim 800$ spectra of $z \sim 3$ LBGs and classified them into 4 categories according to their rest-frame Ly$\alpha$ equivalent widths (EWs). 
They made a composite spectrum of each category and found that LBGs with smaller Ly$\alpha$ EW tend to show larger EWs of low-ionization interstellar (LIS) absorption lines, larger velocity difference between Ly$\alpha$ and LIS absorption, and redder rest-frame UV continua.
Follow-up spectroscopic observations have been made for
LBGs at $z \gtrsim 5$
 (e.g., \citealp{leh03,sta03,sta04,sta07,and04,and07,dow05,dow07}). 
As redshift increases, the targets become fainter and
the characteristic spectral features move into the wavelength region 
where night sky emissions are severe, thus detailed spectroscopic studies of
$z \gtrsim 5$ LBGs are still not easy. 
Hence the sample size of spectroscopically identified $z \sim 5$ LBGs is still very 
small.
In addition, the spectroscopic studies have so far been relying on the Ly$\alpha$
emission, and the features seen in the continuum are 
still not clear except for  rare bright objects such as gravitationally 
lensed LBGs \citep{fry02,swi07}.
Thus a larger deep spectroscopic sample of LBGs at  $z \gtrsim 5$ is 
required to reveal spectroscopic properties of LBGs.

We have constructed a large sample of LBGs at $z \sim 5$ based on Subaru/Suprime-Cam observations \citep{iwa03,iwa07}, and we are conducting spectroscopic observations of selected targets from the photometric sample.
The target fields are the region including the GOODS-N and the J0053+1234 region. 
The total area of the survey fields is 1290 arcmin$^2$ and 228 objects are obtained with $z'<25.0$ mag  i.e. $L > L^*$ in the UVLF of $z \sim 5$ LBGs \citep{iwa07}.
Results of our follow-up spectroscopy with Faint Object Camera And
Spectrograph (FOCAS) on Subaru Telescope were reported by
\citet{and04,and07}, and the number of bright ($z'<25.0$ mag) $z \sim 5$
LBGs with spectroscopic identification was nine and that of faint  
($z'\geq 25.0$ mag) LBGs was two.
Combining the data with those from literature, \citet{and06} claimed the deficiency of bright LBGs with large EWs of Ly$\alpha$ emissions at $z \sim 5$ and $z \sim 6$.
However, the sample size of our spectroscopically confirmed LBGs at $z \sim 5$ was still very small.
Thus we intended to increase the size of the spectroscopic sample.
	
In this paper we present the results of spectroscopic observations of $z \sim 5$ LBGs in the region including the GOODS-N and the J0053+1234 region with  Gemini Multi-Object Spectrograph North (GMOS-N) and South (GMOS-S), respectively.
Gemini/GMOS spectrographs have nod-and-shuffle capability, which enables us to subtract sky emission more clearly and helps the detection of continuum features.
In section \ref{observation}, we describe our sample selection, observations, and data reduction. 
The results and obtained spectra are presented in section \ref{results}. 
In section \ref{discussion}, we discuss distributions of redshifts and colors, an implication to the UVLF of LBGs at $z \sim 5$, and rest-frame EWs of Ly$\alpha$ emission, combining present results with previous data by \citet{and04,and07} and by others.
Throughout this paper, we used a flat $\Lambda$ cosmology; $\Omega_{\mathrm{M}} = 0.3$, $\Omega_\Lambda=0.7$, and $H_0 = 70$ km s$^{-1}$ Mpc$^{-1}$. 
All magnitudes are given in the AB system \citep{oke83}.

\section{TARGET SELECTION, OBSERVATIONS, AND DATA REDUCTION}
\label{observation}
The photometric sample of LBGs at $z \sim 5$ was obtained in the region including the GOODS-N and the J0053+1234 region, based on $V$-, $I_{\mathrm{C}}$-, and $z'$-band images taken with Subaru/Suprime-Cam \citep{iwa03,iwa07}. 
The color criteria for  $z \sim 5$ LBGs are
\begin{equation}
	V - I_{\mathrm{C}} > 1.55 \label{criteria1}
\end{equation}
and
\begin{equation}
	V - I_{\mathrm{C}} > 7.0 (I_{\mathrm{C}} - z') + 0.15. \label{criteria2}
\end{equation}
The sample size is $\sim 600$ objects ($z'<26.5$ mag) in the region including
the GOODS-N and $\sim 200$ objects ($z'<25.5$ mag) in the J0053+1234 region.
More details of the imaging observations and the color selection are described by \citet{iwa07}.
We selected bright ($z'<25.0$ mag) $z \sim 5$ LBG candidates as main spectroscopic targets, aiming at detecting the continuum and absorption features.
Because the entire survey field is too wide to obtain the spectra of all LBG candidates in the survey field, we selected multi-object spectroscopy (MOS) fields to cover as many main targets as possible.
When two slits in the mask design were in conflict, 
we chose the slit of the object with higher surface brightness.
We filled the unused parts of the masks with as many faint 
($z'\geq 25$ mag)  targets  as possible.
We designed 3 masks in the GOODS-N, its flanking field, and the J0053+1234 region.
The numbers of bright LBGs in the masks are 7, 5, and 10, respectively, and the numbers of faint objects are 1, 2, and 0, respectively. 
In the J0053+1234 region, we also observed five objects outside, but near the border of, our color selection criteria, in order to examine our color selection criteria.

Preference to the higher surface brightness mentioned above may introduce
a bias to the spectroscopic sample.
Thus we examined average surface brightness within $r_{50}$
(50\% light encircled radius from SExtractor) as well as
concentration parameter ($C= 5$ log $(r_{80}/r_{20}))$
of the  spectroscopic sample among the whole photometric sample 
in the GOODS-N region including the samples by Ando et al. (2004, 2007).
It is found that in $z'<25.0$ mag the spectroscopic sample is fairly
chosen from the whole sample, while in $z'>25.0$ the spectroscopic sample
tends to bias to LBGs with the higher average surface brightness and
the higher concentration.

Optical spectroscopy was made by using the nod-and-shuffle (micro-shuffle) 
mode of GMOS \citep{hoo04} attached to the 8m Gemini Telescope North
and South.
All of the observations were executed in queue mode
during May to June 2007 for the GOODS-N region 
(Gemini programs GN2007A-Q018-02, GN2007A-Q018-04) and from October 2007 to January
2008 for the J0053+1234 region (Gemini program GS2007B-Q206-01).
We used the R400 grating blazed at 7640 \AA\  with the order cut filter of OG515.
Slits of 1$^{\prime\prime}$ width and
$4^{\prime\prime}-9^{\prime\prime}$ lengths were used and the spectral resolution was $\sim 8$ \AA\ measured from night sky emission lines.
Each spectrum covers the wavelength range of $ \sim 5500-10000$\AA{} depending on the slit position on the mask. 
The exposure time for individual frame was 1800 s, and total exposure times were 10, 8, and 15.5 hours for the GOODS-N, its flanking field, and the J0053+1234 region, respectively. 
Nod interval for each position was  $45$ s for all observations, and the nod distances were $3^{\prime\prime}$, $3^{\prime\prime}$, and $1.4^{\prime\prime}$ for the GOODS-N, its flanking field, and the J0053+1234 region, respectively.
Seeing sizes were typically about 0.7$^{\prime\prime}$ for both GMOS-N and GMOS-S.
Details of observations for each field are listed in Table \ref{obs1} and Table \ref{obs2}. 

The data were reduced with the Gemini IRAF\footnote[7]{Image Reduction and Analysis Facility, distributed by National Optical Astronomical Observatories (NOAO), which are operated by the Association of Universities for Research in Astronomy (AURA), Inc., under cooperative agreement with the National Science Foundation.} package, standard IRAF
 packages, and custom code using the FITSIO package. 
First, bias subtraction was made with combined bias frame. 
Next we shifted the images with the shuffle distance along the slit and subtracted them from the images before shifting to remove the night-sky emissions. 
The resultant images were flat-fielded using the dome flat images taken at the time closest to the time when the object images were taken. 
Then the images were combined after correcting a small offset of the spectra in each 
exposure.
Wavelength calibration was made by using the night sky emission lines. 
An accuracy estimated from night sky emission lines was $\sim 0.3-0.7$\AA.
One dimensional positive and negative spectra were extracted with our custom code, with the aperture determined by eye. 
We combined the positive and negative spectra and applied sensitivity
correction to them using the spectra of standard stars (Feige 66 for
GMOS-N and LTT 9239 for GMOS-S).
We did not make a flux calibration.
The final spectra were obtained by binning the pixels along the wavelength direction to improve the S/N.
The numbers of pixels binned are 5 and 10 for spectra obtained by GMOS-N
and GMOS-S, respectively, to have an identical wavelength bin size ($13.4$\AA).

\section{RESULTS}
\label{results}

\subsection{LBGs Identified to be $z\sim5$}
\label{redshift determination}
Among 22 bright LBG candidates, we identified four objects to be LBGs at $z \sim 5$.
We also identified one object as a $z = 5.15$ LBG among 
three faint targets based on its Ly$\alpha$ emission.  
The resultant spectra of the LBGs identified are shown 
in Figure \ref{spectrum}.

N106944 shows a clear continuum depression and some LIS absorption lines (Si{\scriptsize {} II} $\lambda$1260, O{\scriptsize {} I}+Si{\scriptsize {} II} $\lambda$1303, and C{\scriptsize {} II} $\lambda$1335), hence we can securely conclude that it is at $z \sim 5$.
The redshift determined from the LIS absorption lines is $4.64\pm0.01$.
N127245, S101900, and S103759 show a single emission line and a
continuum depression
in a wavelength region shortward of the emission line, hence their identifications are also secure.
The redshifts of N127245, S101900, and S103759 determined from the Ly$\alpha$ emission line are $4.42\pm0.01$, $4.61 \pm 0.01$, and $4.82 \pm 0.01$, respectively\footnote[8]{Two identified LBGs in the J0053+1234 region, S101900 ($z=4.61$) and S103759 ($z=4.82$), had been previously identified by \citet{ste99} in their $z\sim4$ survey, where they were found to be at the upper end of their redshift selection window.}.
The other identified LBG, N141368 that is chosen from the faint LBG
sample ($z' =25.32$ mag),  shows a single emission line at 7478\AA.
If the emission line is [O{\scriptsize{} III}]$\lambda$5007 or H$\beta$, its redshift is
0.49 or 0.54, respectively.
In this case, a strong  H$\alpha$ emission line is expected to come to
9803{\AA} or 10097\AA, respectively.
However, no significant emission is seen at this wavelength, though the signal-to-noise ratio (S/N) in such a red region is very low.
If the emission line is [O{\scriptsize {} III}]$\lambda$5007, the [O{\scriptsize {} III}]$\lambda$4959 should be seen at 7406{\AA} with a 1/3 flux, which is not seen.
Another possibility is an identification of [O{\scriptsize {} II}]$\lambda$3727.
If this is the  case, an H$\beta$ emission line should come to 9754\AA.
Again no significant emission line is seen.
Since the continuum feature longward  of the emission line is slightly seen in the wavelength regions where the sky emission is weak and there seems to be a  break around at 7460\AA, we identified this emission line as Ly$\alpha$.
The redshift determined from the Ly$\alpha$ emission is $5.15 \pm 0.01$.
The properties  of the LBGs identified are summarized in Table \ref{lbg2}. 

Among 18 remaining bright LBG candidates, we concluded one object (N111905) is a foreground contamination as described in section \ref{foregroundobjects}.
The remaining objects were not identified because of low S/N in their spectra or no spectral feature in the observed wavelength coverage.
Combining with our previous results by \citet{and04,and07}, the total number of our spectroscopic LBG sample is 16.

We examined the identification rates of the spectroscopic sample
including the results by \citet{and04,and07}.
Targets with the higher average surface brightness tend to be identified 
efficiently; in $z'<25$ mag, among targets with average surface brightness 
higher than 29.8 mag arcsec$^{-2}$ the identification rate is $\sim 60$\%, 
while it is $\sim 10$\% among targets with fainter average surface brightness.
Among targets with $z'<25$ mag, highly concentrated objects tend to be 
identified; $\sim 50$\% are identified among objects with $C<2.5$, 
while only $\sim 20$\% for those with $C>2.5$.
These trends, however, can not be seen among targets with $z'>25$ mag;
identification rate is higher in more extended objects, though the subsample
size is smaller.
It should be worth mentioning that the identified LBGs with $z'>25$ mag
show strong Ly$\alpha$ emission, while most of the identified LBGS with
$z'<25$ mag show no or very weak Ly$\alpha$ emission.
The trend that luminous LBGs do not show strong Ly$\alpha$ emission prevents
us from achieving a high identification rate even in bright LBG sample.
The trend and its physical cause will be discussed in $\S 4.3$.

\subsection{Foreground Objects}
\label{foregroundobjects}
We identified three foreground objects; one is in our color criteria and the others are outside of the criteria.
The former (N111905) shows an emission line at 8209\AA{} and the continuum is seen shortward of the emission line.
Thus this object should be a foreground object.
If the emission line at 8209\AA\  is [O{\scriptsize {} II}]$\lambda$3727, its redshift is $1.20$, and no other major emission lines seen in star-forming galaxies come into the observed wavelength coverage. 
If the emission line is one of H$\beta$, [O{\scriptsize {} III}]$\lambda$4959, or [O{\scriptsize {} III}]$\lambda$5007, other lines should
appear around the emission line, but no other emission lines can be seen.
Hence probably the object is a foreground object at $z=1.20$.
However, its $V-I_{\mathrm{C}}$ color is somewhat redder than the expected color for a star-forming galaxy at $z=1.20$.
It is worth noting that the object is located quite close to the selection criteria (see Section \ref{redshiftdistributionofLBGsanditsrelationwithphotometriccolors}). 
The contamination rate in our spectroscopic sample with $z'$-band magnitude from $24.0$ to $25.0$ in the GOODS-N region is 1/9 (including those from \citealp{and04,and07}), which is not larger than the estimated value of 17 \% \citep{iwa07}.

One of the five targets outside of the color selection window is identified to be a Galactic M star. 
Its $V-I_{\mathrm{C}}$ and $I_{\mathrm{C}}-z'$ colors are consistent with typical colors of Galactic M stars (see Section \ref{redshiftdistributionofLBGsanditsrelationwithphotometriccolors}).
Another object outside of the color selection window is identified to be an elliptical galaxy at $z =0.39$.
Its spectrum clearly shows a Mgb absorption, 4000\AA\ break, and Ca H and K absorptions.
No emission lines are seen.
However, its $V-I_{\mathrm{C}}$ and $I_{\mathrm{C}}-z'$ colors are somewhat bluer than the expected colors for an early-type galaxy at $z=0.39$.
The remaining objects were not identified because of low S/N in their spectra or no spectral feature in the observed wavelength coverage.

%These results ensure the robustness of the color-selection window by \citet{iwa03,iwa07}.

\section{DISCUSSION}
\label{discussion}

\subsection{Redshift and  Color Distribution of $z\sim5$ LBGs}
\label{redshiftdistributionofLBGsanditsrelationwithphotometriccolors}
\begin{figure}[!t]
\begin{center}
\epsscale{0.8}
\plotone{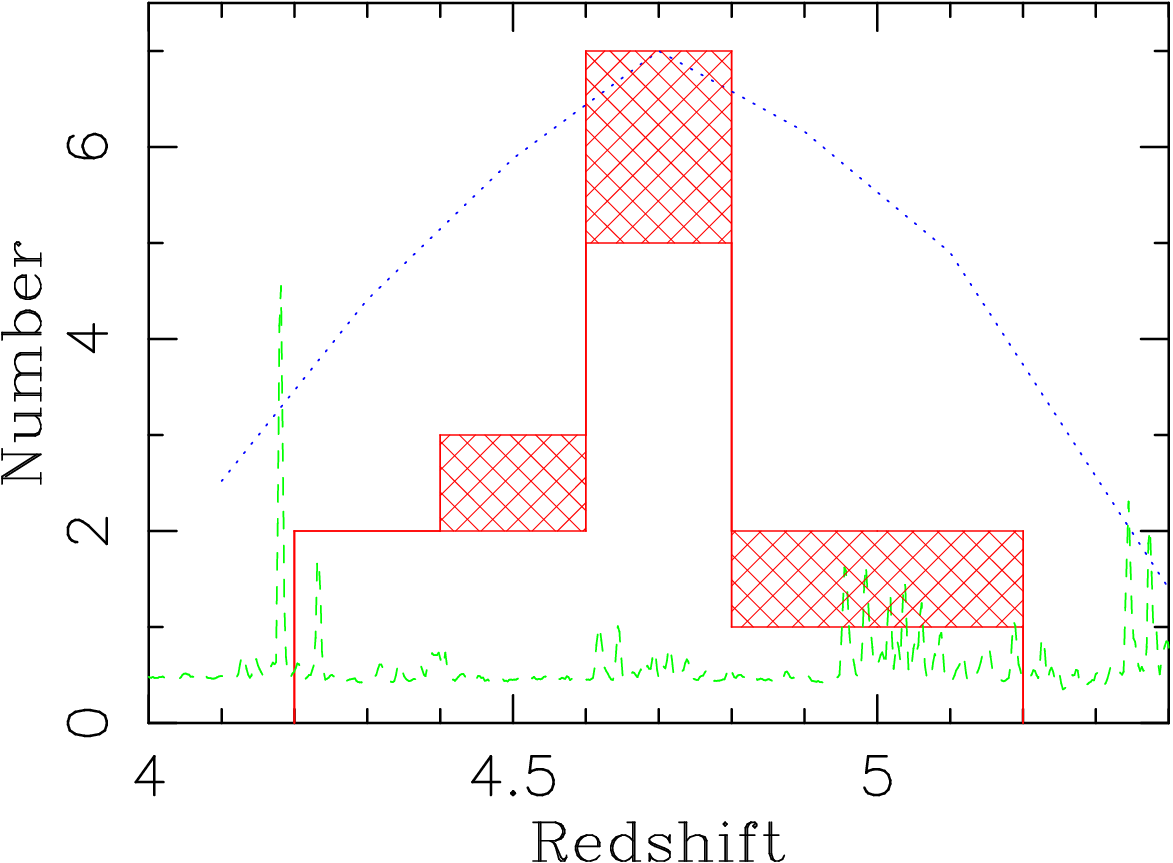}
\caption{
	Redshift distribution of identified LBGs.
	The hatched histogram shows the LBGs identified in this study, and the white histogram shows the LBGs from previous studies by \citet{and04,and07}.
	The dotted line is the normalized expected redshift distribution. The dashed line represents the sky intensity at the redshifted Ly$\alpha$ wavelength.
}
\label{redshift}
\end{center}
\end{figure}

Figure \ref{redshift} shows the redshift distribution of spectroscopically confirmed $z \sim 5$ LBGs in this study and that by \citet{and04,and07}.
Hatched and white histograms show $z \sim 5$ LBGs from the present
results and from the  FOCAS sample by \citet{and04,and07}, respectively. 
We show the expected redshift distribution of our sample of $z \sim 5$ LBGs as a dotted line normalized at $z = 4.7$. 
The expected distribution was calculated based on the detection rates of the LBGs against the apparent magnitude and the redshift for each survey field by considering the survey volume and reached depth in UVLF at each redshift bin \citep{and07}. 
We also plotted the intensity of night sky emission at the wavelength corresponding to the redshifted Ly$\alpha$. 
Although the observed redshift distribution seems to be concentrated at $z=4.7$, it is consistent with the expected distribution within the statistical uncertainty.
	
\begin{figure}[!t]
\begin{center}
\epsscale{0.5}
\plotone{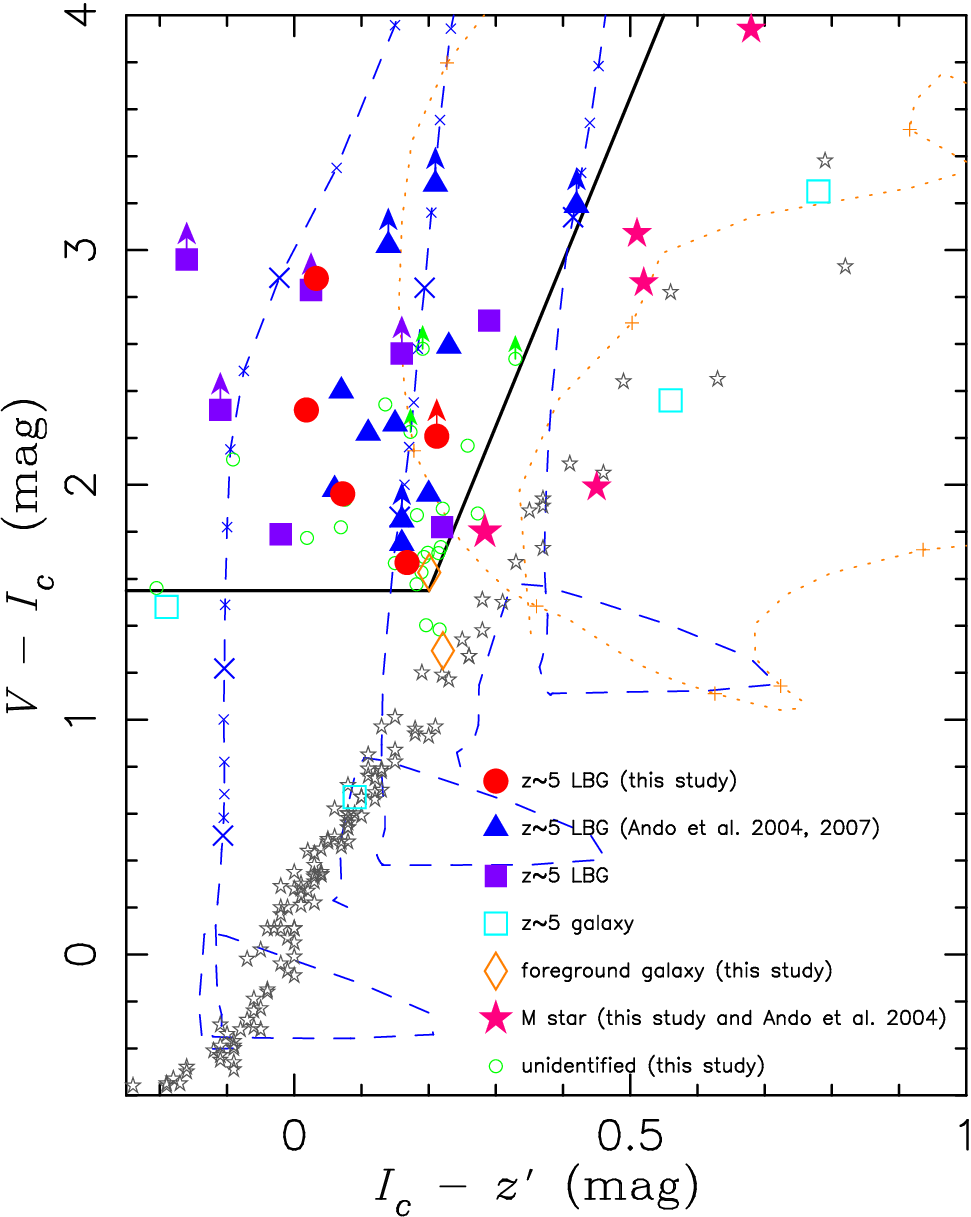}
\caption{{\small
	$V-I_{\mathrm{C}}$ vs $I_{\mathrm{C}}-z'$ diagram.
	Filled circles show spectroscopically identified LBGs in this study.
	Filled triangles show the LBGs by \citet{and04,and07}.
	Open diamonds show the foreground galaxies found in this study.
	Filled stars show the Galactic stars in this study and those by \citet{and04}.
	Small open circles represent LBG candidates unidentified in this study.
	Filled squares refer to LBGs  which are detected by \citet{iwa07} and spectroscopically identified by other studies \citep{spi98,ste99,daw01,daw02,fer01,bar08}.
	Open squares refer to spectroscopically identified $z\sim 5$ galaxies \citep{ster99,daw01,cow04}, but not in the color criteria.
	The solid line show our selection color criteria (eq (\ref{criteria1}) and (\ref{criteria2})).
	Dashed lines show the model color tracks of star-forming galaxies from \citet{iwa07}, with $E(B-V) = 0.0$, $0.4$, and $0.8$ mag from left to right, respectively (see text for details).
	Small crosses are plotted on the tracks for $z \ge 4.0$ with a redshift interval of 0.1. 
	Symbols at $z=4.0$, $4.5$, and $5.0$ are enlarged.
	The dotted line represents the color track of an early-type galaxy \citep{col80}.
	Small pluses are plotted on the track with a redshift interval of 0.5. 
	Small open stars represent colors of Galactic A0-M9 stars calculated from the library by \citet{pic98}.
}}
\label{cdiagram}
\end{center}
\end{figure}

Figure \ref{cdiagram} shows the positions of the observed LBGs in the two-color ($V-I_{\mathrm{C}}$ and $I_{\mathrm{C}}-z'$) diagram. 
A solid line shows our color selection criteria represented by eq. (\ref{criteria1}) and (\ref{criteria2}) \citep{iwa07}. 
Filled circles and filled triangles show the LBGs observed in this study and those by \citet{and04,and07}, respectively. 
Open diamonds show the foreground galaxies found in this study.
%As mentioned in section \ref{foregroundobjects}, one foreground galaxy
%resides in our selection criteria, 
%but it is very close to the boundary of the color window. 
Filled stars show the Galactic stars in this study and those by \citet{and04}.
Small open circles show LBG candidates unidentified in this study.
Three dashed tracks show model colors of star-forming galaxies.
The model spectrum of a star-forming galaxy is calculated with P\`{E}GASE
version 2 \citep{fio97}, by assuming constant star-formation history
with an age of $100$ Myr, Salpeter initial mass function (IMF;
\citealp{sal55}) with upper mass of 120 $M_{\odot}$ and lower mass
of 0.1 $M_{\odot}$, and the dust extinction curve by \citet{cal00}. 
Average IGM attenuation is calculated based on a prescription by
\citet{ino05}.
Three tracks represent models of $E(B-V)$ of 
0.0, 0.4, and  0.8 mag. 
A dotted line represents a color track of an early-type galaxy \citep{col80}.
 As mentioned in section \ref{foregroundobjects}, one foreground galaxy
resides in the color selection window, 
but it is very close to a boundary of the window.
While majority of the spectroscopically confirmed $z \sim 5$ 
LBGs lie away from the boundary,  unidentified objects are close to it.

In Figure \ref{cdiagram}, $z\sim5$ galaxies spectroscopically 
identified by other studies in the survey fields 
(Tables 3 and 4 by \citealp{iwa07}; \citealp{bar08}) are also plotted .
Filled and open squares show the identified objects in the
color-selection window and those outside of it, respectively.
Properties of all spectroscopically identified objects at $z \ge 4.2$ are listed in Tables \ref{speclbg1} and \ref{speclbg2}.
Among them, one object shows peculiar colors of  $I_{\mathrm{C}}-z' \sim 0.1$ mag 
and $V-I_{\mathrm{C}} \sim 0.7$ mag. 
\citet{iwa07} suggest that this object may be a foreground object 
because the original identification was based on a single emission line.
Other two objects with red $I_{\mathrm{C}}-z'$ colors may be dust reddened 
star-forming galaxies.
However, if the redness of these objects is really due to the dust extinction,
the extinction at rest-frame 1600 \AA\ corresponds to 8 mag or more and the intrinsic luminosity is huge, if we assume the
Calzetti extinction curve.
In addition, their $V-I_{\mathrm{C}}$  colors are not consistent with their redshifts;
$V-I_{\mathrm{C}}$ colors are too blue to be at $z\sim5$. 
These  may imply that the spectral models we adopt do not
cover the whole real spectra of $z\sim5$ galaxies. 
Since we did not take spectra of objects outside of the color criteria 
extensively, further studies on such galaxies are 
desirable to know their nature and fraction.

\subsection{Implications to the UV Luminosity Function of LBGs at $z \sim 5$}
\label{LF}
Using the photometric sample, \citet{iwa07} derived the UV luminosity function (UVLF) of LBGs at $z \sim 5$ in the region including the GOODS-N and the J0053+1234 region.
They found that there is a significant population of bright 
($M_{UV} < -22.0$ mag) LBGs at $z \sim 5$ comparable to that of 
$z \sim 3-4$, while the faint ($M_{UV} > -21.0$ mag) 
end of their UVLF shows a gradual increase from $z \sim 5$ to $z \sim 3$ 
\citep{saw06}.
So \citet{iwa07} suggest luminosity dependent 
evolution of LBGs at these redshifts.
However, different results are derived from other studies of UVLF of 
LBGs at $z \sim 5$.
The UVLFs in Subaru Deep Field and Subaru XMM-Newton Deep Field derived 
by \citet{ouc04a} and by \citet{yos06} show a smaller 
number density of bright LBGs than that found by \citet{iwa03,iwa07}, and
suggest an evolution of the number density in the bright part of the UVLF
from $z \sim 5$ to $z \sim 3$.
The UVLFs by \citet{bec06} and \citet{bou07} also show a
 similar trend to those by \citet{ouc04a} and \citet{yos06}.
The difference between the number density of bright ($M_{UV} < -22.0$
mag) LBGs of \citet{iwa03,iwa07} and that of \citet{ouc04a} and \citet{yos06}
is $0.5-1.0$ dex.

The cause of the divergence of UVLFs is still unknown.
Field-to-field variance may exist.
Or different filter sets used in various LBG surveys 
may cause the difference \citep{sta08a}.
The spectroscopic sample helps to constrain the UVLF.
We derived the lower limit of the number density of LBGs 
at $z \sim 5$ in the GOODS-N region and the J0053+1234 region, using the 
spectroscopically confirmed LBGs including the spectroscopy 
from the literature \citep{daw01,daw02,fer01,spi98,ste99,bar08},
but not including $z\sim5$ galaxies outside of the color selection window.
We derived the lower limits for each field by dividing the numbers of
spectroscopically confirmed LBGs  by the effective volume in
each magnitude bin from \citet{iwa07}.
Then we averaged the lower limits of two fields weighting with
their survey areas.
When we calculate the UV absolute magnitude of the 
spectroscopic sample, we used the fixed redshift of $z = 4.8$ 
for consistency with the estimation of the effective volume.
The difference of the UV absolute magnitude by this assumption and that
from spectroscopic redshift is $\lesssim 0.2$ mag.

\begin{figure}[!t]
\begin{center}
\epsscale{0.7}
\plotone{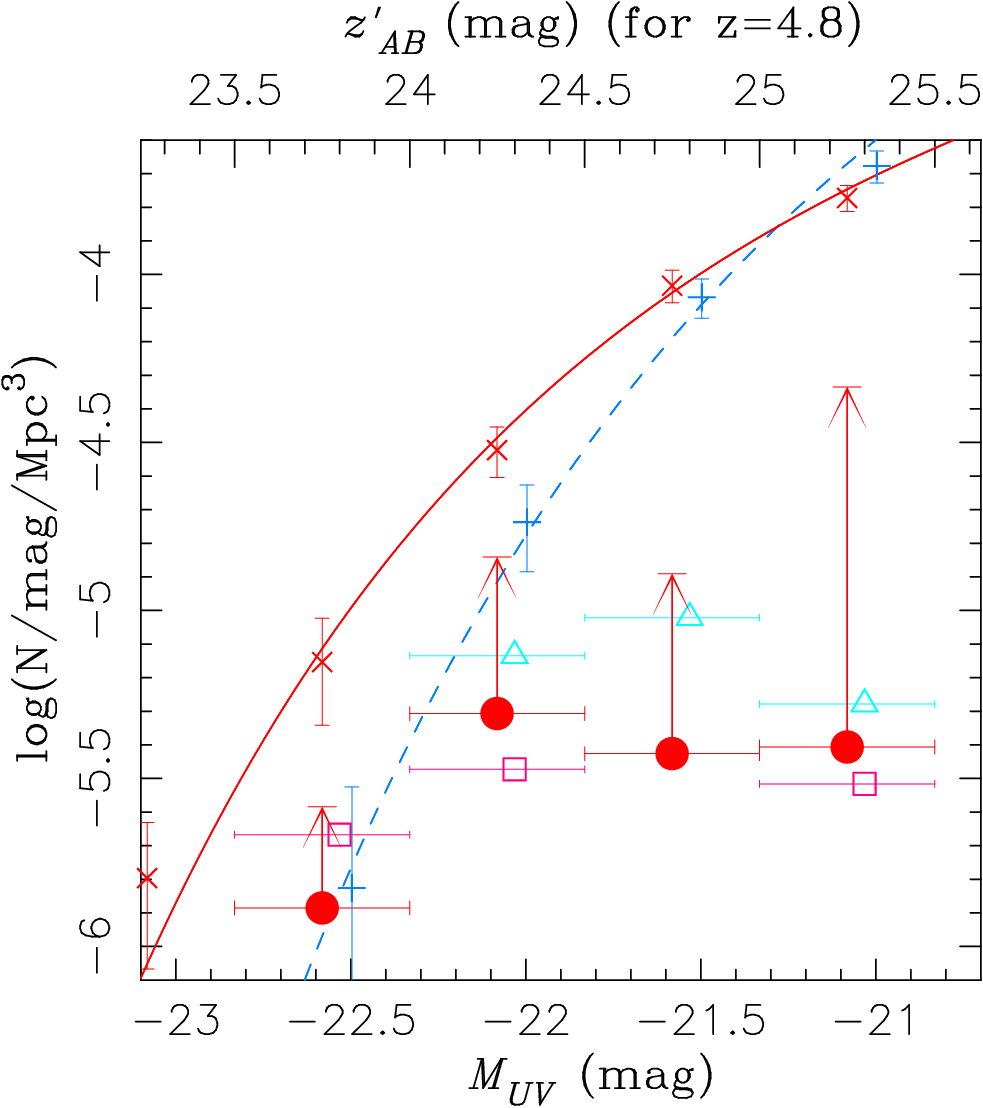}
\caption{
	Lower limits on the number density of LBGs at $z \sim 5$.
	Open triangles, open squares, and filled circles show the lower limits on UVLF in the GOODS-N region, the J0053+1234 region, and the average of them, respectively.
	The data points of lower limits in each field are shifted +$0.05$ mag for clarity.
	Vertical arrows show the expected values derived by multiplying the number of photometric sample by the success rate of the spectroscopic sample in this study and that by \citet{and04,and07}.
	Note that unidentified objects are not included in the lower limits.
	The solid and the dashed line shows the Schechter function fit to the UVLF by \citet{iwa07} and \citet{yos06}, respectively with the data points (small crosses and pluses, respectively).
}
\label{UVLF}
\end{center}
\end{figure}

Figure \ref{UVLF} shows the derived lower limits on the number density 
of LBGs at $z \sim 5$ in the GOODS-N region, the J0053+1234 region, and their average.
The solid and the  dashed line shows the Schechter function fit to the UVLF
 of \cite{iwa07} and \cite{yos06}, respectively with the data points (small crosses and pluses, respectively).
The conservative estimation of the lower limits by using only the spectroscopic sample gives the number density of LBGs comparable to that by \citet{yos06} in the magnitude range of $23.5 < z' < 24.0$ mag, and a slightly smaller value in $24.0 < z' < 24.5$ mag.
The fractions of spectroscopically confirmed LBGs to
the total photometric sample in the magnitude range 
of $23.5 < z' < 24.0$ mag, $24.0 < z' < 24.5$ mag,
$24.5 < z' < 25.0$ mag, and $25.0 < z' < 25.5$ mag are 
0/5, 5/35, 6/72, and 3/126, respectively in
 the GOODS-N, and 2/10, 3/22, 0/79, and 2/122, respectively in the J0053+1234 region. 
Considering that more than half of the photometric sample by 
\citet{iwa07} are not observed yet, the number density of LBGs 
in each magnitude bin is expected to be larger than this lower limit
by a factor two or more, and to approach toward the values by \citet{iwa07}.
We estimated the expected number densities if the full photometric sample were spectroscopically observed, by multiplying the number of the photometric sample by success rate of spectroscopy in each magnitude bin, and by dividing it by the effective volume from \citet{iwa07}.
Results are shown with arrows in Figure \ref{UVLF}.
The expected number densities of bright ($M_{UV} < -22.0$ mag) LBGs are comparable to or slightly larger than those by \citet{yos06}.
Since we regarded all unidentified objects as foreground objects in this estimation, true number densities are probably larger than these estimations.
Thus the number densities in the bright part from spectroscopic sample are still consistent with those from \citet{iwa03,iwa07}.

\subsection{Rest-frame UV Luminosity and Ly$\alpha$ EW}
\label{Rest UV Luminosity and Lyalpha EW}
\begin{figure}[!t]
\begin{center}
\epsscale{0.6}
\plotone{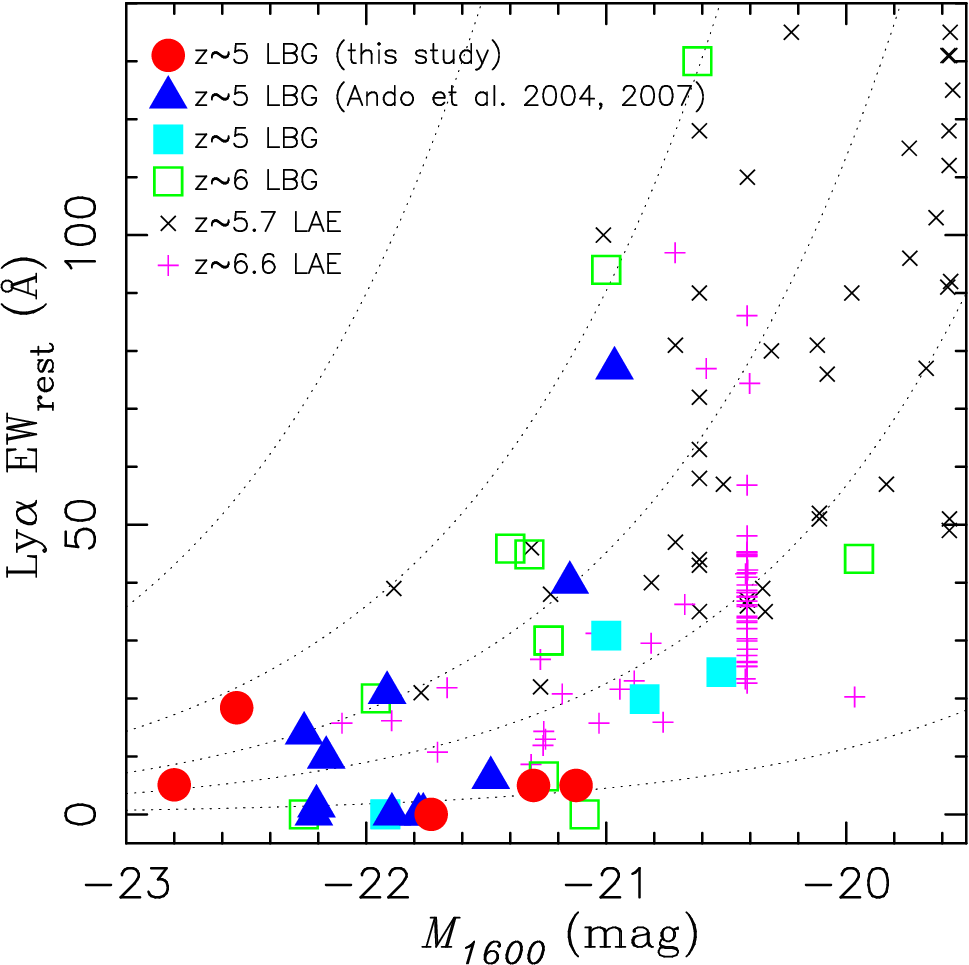}
\caption{
	Rest-frame equivalent widths (EWs) of Ly$\alpha$ emission against absolute magnitude at rest-frame 1600\AA{} for galaxies at $z \sim 5$ and $z \sim 6$.
	Filled circles and filled triangles show our spectroscopic $z \sim 5$ LBGs in this study and those by \citet{and04,and07}, respectively.
	The other symbols show LBGs and LAEs from literature.
	Filled squares, open squares, filled triangles, and open triangles represent $z \sim 5$ LBGs \citep{leh03,sta07}, $z \sim 6$ LBGs \citep{leh03,sta03,sta04,sta07,nag04,nag05,dow05,dow07}, $z \sim 5.7$ LAEs \citep{aji03,shi06}, and $z \sim 6.6$ LAEs \citep{tan05}, respectively. 
	We used the EWs spectroscopically derived for a part of sample of \citet{tan05}.
	Dotted lines show constant Ly$\alpha$ luminosities of $5 \times 10^{43}$, $2 \times 10^{43}$, $10^{43}$, $5 \times 10^{42}$, and $10^{42}$ ergs s$^{-1}$ from top-left to bottom-right.
}
\label{MUVvsEW}
\end{center}
\end{figure}

\citet{and06} reported the absence  of rest-frame UV luminous LBGs with large EW of Ly$\alpha$ emission at $z \sim 5$. 
Combining recent results by \citet{sta07} and \citet{dow07}, we update the plot of rest-frame Ly$\alpha$ EW against rest-frame UV absolute magnitude and show it in Figure \ref{MUVvsEW}.
Rest-frame UV absolute magnitudes of the LBGs and LAEs were calculated from their $z'$-band magnitudes. 
We chose UV absolute magnitude at 1600\AA\ ($M_{1600}$) to minimize the effect of uncertainty of continuum slope. 
We selected $\beta = -2$ ($f_\lambda \propto \lambda^{\beta}$) as a  typical value 
for young starburst.
The value of $\beta = -2$ is  consistent with the observed $I_{\mathrm{C}}-z'$ colors of spectroscopically confirmed LBGs. 
The uncertainty of $M_{1600}$ due to the assumption of $\beta$ ($\beta = -2 \pm 1$) is about $0.1-0.2$ mag.  
If the continuum flux densities at longward of Ly$\alpha$ line are listed 
in the literature, we used them to derive the rest-frame UV absolute 
magnitudes to mitigate the influence by strong Ly$\alpha$ emission and
IGM attenuation in the band (\citealp{nag04,nag05,tan05}). 
Filled circles and filled triangles show spectroscopically confirmed $z \sim 5$ LBGs in this study and that by \citet{and04,and07}, respectively.
Other symbols show LBGs and LAEs from the literature.
Filled squares, open squares, crosses, and pluses represent $z \sim 5$ LBGs \citep{leh03,sta07}, $z \sim 6$ LBGs \citep{leh03,sta03,sta04,sta07,nag04,nag05,dow05,dow07}, $z \sim 5.7$ LAEs \citep{aji03,shi06}, and $z \sim 6.6$ LAEs \citep{tan05}, respectively.

The deficiency of UV luminous LBGs with large Ly$\alpha$ EW is seen 
in our revised plot. 
There are no luminous ($M_{1600}<-21.5$ mag) $z \sim 5$ LBGs 
with large ($>20$\AA) EW Ly$\alpha$ emission.
Although we may miss the faint LBGs with small Ly$\alpha$ EW, 
we can detect luminous LBGs  with large Ly$\alpha$ EW.
Thus  the deficiency must be real.
Such distribution is also seen among $z \sim 6$ LBGs, $z \sim 5.7$ LAEs, 
and $z \sim 6.6$ LAEs. 
The threshold UV magnitude seems to lie around $-21.5$ mag $ < M_{1600} < -21.0$ mag, and this value is close to $M_*$  of our $z \sim 5$ LBG sample ($M_* = -21.28$ mag; \citealp{iwa07}). 
A similar trend can be seen among $z\sim3$ LBGs \citep{sha03};
a composite spectrum of UV luminous LBGs shows smaller Ly$\alpha$ EW than
that of UV faint ones.

The deficiency of UV luminous LBGs with large Ly$\alpha$ EW
may reflect the deficiency of LBGs (and LAEs) with large Ly$\alpha$ luminosity.
Dotted lines in Figure \ref{MUVvsEW} represent constant  Ly$\alpha$ luminosities of $5 \times 10^{43}$, $2 \times 10^{43}$, $10^{43}$, $5 \times 10^{42}$ and $10^{42}$ ergs s$^{-1}$ from top-left to bottom-right.
It appears that very few objects show Ly$\alpha$ luminosity larger than $2 \times 10^{43}$ ergs s$^{-1}$.
Since the number density of galaxies with such large Ly$\alpha$ 
luminosity is small, the deficiency may be due to poor 
statistics (e.g., Nilsson et al. 2009).
Meanwhile, galaxies under constant star formation with ages of
$10-100$ Myr are expected to show  EWs of $100-200$ \AA\ intrinsically.
(We used the P\`{E}GASE version 2 \citep{fio97}, with Salpeter IMF
\citep{sal55} with an upper-mass limit of 120 $M_{\odot}$, and assumed
the case B recombination and all Lyman continuum photons ionize neutral 
hydrogen.)
Thus, the  deficiency of large Ly$\alpha$ EWs among UV luminous 
($M_{1600}<-21.5$) galaxies is still mysterious.
Possible causes of small EWs in UV luminous LBGs 
(except for the poor statistics)
are (i) difference of escape fraction of Ly$\alpha$ emission; Ly$\alpha$ emission is selectively quenched by dust extinction and/or scattering by neutral hydrogen in more UV luminous galaxies, (ii) difference of time scale of star formation; the UV continuum is the probe of star-formation with longer time scale than the nebular emission lines, and star-formation age is larger in more UV luminous galaxies, and (iii) difference of IMF in the galaxies; deficiency of massive stars in more UV luminous galaxies.

The escape of Ly$\alpha$ emission is a complex problem, and it is thought to depend on the geometry, dynamics, column density, and dust content of neutral hydrogen in/around the star-forming regions.  
In nearby starburst galaxies, the existence or absence of outflow of
neutral hydrogen largely affects the escape fraction, and the effect of
dust content is small (e.g. \citealp{kun98,mas03,ate08}). 
These authors showed that starburst galaxies with static interstellar media show Ly$\alpha$ as absorption, whereas galaxies showing velocity offset between interstellar absorption lines and Ly$\alpha$ show Ly$\alpha$ as emission lines.
In high redshift galaxies, on the other hand, $z \sim 3$ LBGs with
larger velocity offset between interstellar absorption and Ly$\alpha$
tend to show smaller Ly$\alpha$ EW \citep{sha03}.
In addition, $z\sim3$ LBGs with a large Ly$\alpha$ EW 
show bluer UV continua  \citep{sha03,tap07},
suggesting that the dust extinction is the main cause for the 
Ly$\alpha$ quench.
At $z\sim5$, although the correlation between the EW and $I_{\mathrm{C}}-z'$ color 
is not clear,
the LBGs with large Ly$\alpha$ EW ($>40$\AA) in our sample and 
that by \citet{and04,and07} (combined sample) show relatively 
blue observed $I_{\mathrm{C}}-z'$ colors ($I_{\mathrm{C}}-z' \sim 0.06$ mag and $0.16$ mag).
The LBG with the largest EW in the combined sample shows a
weak or no LIS absorption lines in its UV continuum.
We roughly estimated the metallicity of nine $z\sim5$ LBGs which show LIS absorption lines
in the combined sample, 
using the empirical relation between the metallicity and the 
EWs of LIS absorption lines \citep{hec98}.
The average value is $12+\log{\rm [O/H]} = 7.7$ ($\sim 1/10$Z$_{\odot}$, 
 \citealp{all01}), suggesting that LBGs at $z \sim 5$ are chemically 
evolved to some extent and expected to contain dust.
These prefer the dust extinction as the major cause 
for the Ly$\alpha$ quench.
If more UV luminous LBGs are more chemically evolved than UV faint ones on average, it suggests that UV luminous LBGs formed earlier \citep{and06,iwa07}. 
It should be mentioned, however,  that color excesses of a part of 
the photometric LBG sample \citep{iwa07} are derived by SED 
fitting by covering rest-frame 
UV to optical wavelength (\citet{yab09}; they assumed the constant star
formation history, the Calzetti extinction curve, and 0.2 solar
abundance as a fiducial model), 
and no clear correlation is 
seen between the derived color excesses and Ly$\alpha$ EWs.
However, the number of LBGs for which both color excesses and spectra are obtained is only five, and the range of the EW 
is very limited (up to only $\sim$ 5\AA).
We further examined a relation between the $z'$ magnitude and
the color excess obtained by SED fitting by \citet{yab09}, 
but found no clear trend.

The age of a galaxy also affects the EW of Ly$\alpha$. 
Young 
starbursts are expected to show  large Ly$\alpha$ EW,
while old galaxies show small EW.
Ages of a part of our LBG sample are also obtained by SED fitting 
\citep{yab09}; no clear relation between the age and the Ly$\alpha$ EW can be seen (the sample size is five as mentioned above). 
We further examined the distribution of ages against $z'$ magnitude by using
results of the SED fitting  \citep{yab09}.
The derived ages range from a few Myr to a few 100 Myr 
with a median of 25 Myr under constant star formation, and 
an absence of young ages ($<10$ Myr) is seen among bright 
LBGs ($z' < 25$ mag).
However, with the supposed IMF (Salpeter IMF with an upper-mass limit
of $100 M_{\odot}$) in the SED fitting, the Ly$\alpha$ EW is 
$\sim 100$\AA\ at ages from 10 Myr to 1 Gyr,  
thus the age is unlikely to be a primary
cause for the small Ly$\alpha$ EW.
If we assume an exponentially  decaying star formation history,
the EW decreases more rapidly.
Since Yabe et al. (2009) also studied the exponentially decaying models with $\tau = 1$ Myr,
10 Myr, 100 Myr, and 1 Gyr, we examined the
cases using their data and found that the UV bright LBGs show
relatively young ages of 1 Myr to 10 Myr with a median $\tau$ of $\sim
1$ Myr.
With these ages, the exponentially decaying model with $\tau=1$ Myr shows
Ly$\alpha$ EWs larger than $\sim 20$ \AA, again suggesting 
that the age is not the primary cause for the deficiency of large Ly$\alpha$
EWs.
It is worth noting here that for the $z \sim 3$ LBGs \citet{sha03} 
claim that the composite spectrum of old LBGs ($>$ 1 Gyr) 
shows larger EW of Ly$\alpha$ emission than young LBGs ($<35$ Myr).
\citet{sha03} interpret  that dust extinction in old LBGs 
is smaller than in young LBGs that are under active star formation.

A top-heavy IMF produces a large Ly$\alpha$ EW; 
in less UV luminous $z\sim5$ galaxies this may occur especially in LAEs.
The Salpeter IMF with an upper-mass limit of 120 $M_{\odot}$
produces more than $100$\AA\  of Ly$\alpha$ EW in a young phase 
($10-100$ Myr) under constant star formation and Case B assumptions as described before.
To reduce the EW in the young phase to less than $20-40$\AA{} by changing an upper-mass limit, 
it should be smaller.
If the upper-mass limit is  $14-15 M_{\odot}$,
the EW is smaller than $\sim 20$ \AA\ at ages larger than 1 Myr.
Although this seems to be unlikely for most UV-luminous class
active star-forming  galaxies at high redshifts,
we cannot rule out this possibility completely.

\section{Summary}
\label{summary}
We presented results of optical spectroscopic observations of $z \sim 5$ LBGs.
The observations were made in the GOODS-N and its flanking field with GMOS-N
and in the J0053+1234 region with GMOS-S.
The number of bright ($z'<25.0$ mag) LBG candidates observed is 22
 and the total number of observed LBG candidates is  25.
We identified four LBGs among 22 bright targets (2 LBGs identified 
in the J0053+1234 region were already spectroscopically identified by
\citealp{ste99}) and one LBG among three faint targets.
The redshift distribution of identified LBGs is consistent with 
the expected redshift distribution.
One target turned out to be a foreground galaxy is located 
almost at  the boundary of the color selection window in 
the two-color diagram.
The foreground contamination fraction is consistent with
model expectations based on simulations \citep{iwa07}.
At the same time two objects just outside of the color selection window
are identified to be foreground objects; one is an M-type star, and the other is an elliptical galaxy at $z = 0.39$. 

Using the full spectroscopically confirmed sample, we derived the raw lower 
limits on the number density of bright ($M_{UV}<-22.0$ mag) LBGs 
at $z \sim 5$.
The lower limits are comparable to or slightly smaller than the number densities 
derived by \citet{yos06}.
However, considering that there remains a number of LBG candidates 
without spectroscopic observations, the number density of
spectroscopically derived bright LBGs is expected to increase 
by a factor of two or more.

We confirmed the deficiency of rest-UV luminous ($M_{1600}<-21.5$) LBGs 
with large rest-frame EW ($>20$\AA) of Ly$\alpha$ emission \citep{and06}.
We discussed possible causes for the deficiency and prefer 
the interpretation of dust absorption rather than gas outflow, age
difference, and IMF difference.

\vspace{5mm}

We would like to thank staff members at Gemini observatory 
for carrying out our observations.
Especially, we are grateful to support scientists Atsuko Nitta 
and Bryan Miller for their helpful comments during preparation.
This work was partly made based on Gemini-Subaru machine time 
exchange program.
This work is supported by Grant-in-Aid for Scientific Research on
Priority Area (19047003) from Ministry of Education, Culture, Sports,
Science, and Technology of Japan, and in part supported 
by funding from the Natural Sciences and Engineering
Research Council of Canada and by the Canadian Space Agency.

\begin{deluxetable}{ccccccc}
\tabletypesize{\scriptsize}
\tablecaption{Journal of observations\label{obs1}}
\tablewidth{0pt}
\tablecolumns{7}
\tablehead{
	\colhead{Gemini ID} & \colhead{Field} & \colhead{Center of Mask}& \colhead{(J2000.0)} & \colhead{Instrument} & \colhead{Observation} & \colhead{Exposure Time} \\
	\colhead{} & \colhead{} & \colhead{R.A.} & \colhead{Decl.} & \colhead{} & \colhead{Period} & \colhead{(s)} 
}
\startdata
GN2007A-Q018-02 & GOODS-N         & 12:36:20.680 & 62:14:13.00 & GMOS-N & 2007 May 25    & $20 \times 1800$ \\
			   &                 &              &             &        & to 2007 Jun 10 &                  \\
GN2007A-Q018-04 & flanking field  & 12:37:51.830 & 62:08:44.00 & GMOS-N & 2007 Jun 17    & $16 \times 1800$ \\
			   & of GOODS-N      &              &             &        & to 2007 Jun 24 &                  \\
GS2007B-Q206-01 & J0053+1234      & 00:53:26.930 & 12:31:45.00 & GMOS-S & 2007 Oct 5     & $31 \times 1800$ \\
			   &                 &              &             &        & to 2008 Jan 6  &                  \\
\enddata
\end{deluxetable}
\begin{deluxetable}{cccccc}
\tabletypesize{\scriptsize}
\tablecaption{Observing setup\label{obs2}}
\tablewidth{0pt}
\tablecolumns{6}
\tablehead{
	\colhead{Gemini ID} & \colhead{Grating} & \colhead{Filter} & \colhead{Slit Size} & \colhead{Pixel Scale} \\
	\colhead{} & \colhead{} & \colhead{} & \colhead{(arcsec $\times$ arcsec)} & \colhead{\AA\ $\times$ arcsec} 
}
\startdata
GN2007A-Q018-02 & R400+\_G5305 & OG515\_G0306 & $1 \times 8$ & $2.7\times 0.15$  \\
GN2007A-Q018-04 & R400+\_G5305 & OG515\_G0306 & $1 \times 9$ & $2.7\times 0.15$  \\
GS2007B-Q206-01 & R400+\_G5325& OG515\_G0330 & $1 \times 3.8$ & $1.3\times 0.15$  \\
\enddata
\end{deluxetable}

\begin{deluxetable}{cccccccccc}
\tabletypesize{\scriptsize}
\tablecaption{Properties of identified LBGs\label{lbg2}}
\tablewidth{0pt}
\tablecolumns{10}
\tablehead{
	\colhead{ID} & \colhead{R.A.} & \colhead{Decl.} & \colhead{$z'$}  & \colhead{$V-I_{\mathrm{C}}$} & \colhead{$I_{\mathrm{C}}-z'$} & \colhead{$M_{1600}$} & \colhead{Redshift} & \colhead{Ly$\alpha$ EW$_{\textrm {rest}}$} \\
	\colhead{} & \colhead{(J2000.0)} & \colhead{(J2000.0)} & \colhead{(mag)} & \colhead{(mag)}  & \colhead{(mag)} & \colhead{(mag)} & \colhead{} & \colhead{(\AA)}
}
\startdata
N106944 & 12:37:58.12 & 62:09:51.3 & $24.55 \pm 0.05$ & $2.32 \pm 0.10$ & $0.018 \pm 0.045$ & $-21.73 \pm 0.05$ & $4.64 \pm 0.01$ & $0$ \\
N127245 & 12:36:35.49 & 62:13:50.3 & $24.89 \pm 0.06$ & $1.96 \pm 0.11$ & $0.072 \pm 0.065$ & $-21.30 \pm 0.06$ & $4.42 \pm 0.01$ & $5.0^{+5.0}_{-1.9}$ \\
N141368 & 12:36:14.21 & 62:16:43.3 & $25.32 \pm 0.13$ & $>$$2.21$       & $0.212 \pm 0.140$ & $-21.13 \pm 0.13$ & $5.15 \pm 0.01$ & $5.0^{+6.0}_{-2.4}$ \\
S101900 & 00:53:35.57 & 12:31:44.1 & $23.72 \pm 0.02$ & $1.67 \pm 0.02$ & $0.168 \pm 0.014$ & $-22.54 \pm 0.02$ & $4.61 \pm 0.01$ & $18^{+18}_{-8}$ \\
S103759 & 00:53:33.21 & 12:32:07.3 & $23.54 \pm 0.02$ & $2.88 \pm 0.03$ & $0.033 \pm 0.012$ & $-22.80 \pm 0.02$ & $4.82 \pm 0.01$ & $5.1^{+3.4}_{-1.7}$ \\
\enddata
\end{deluxetable}
\begin{deluxetable}{lccccccc}
\tabletypesize{\scriptsize}
\tablecaption{Spectroscopically identified LBGs in the sample by \citet{iwa07}\label{speclbg1}}
\tablewidth{0pt}
\tablecolumns{8}
\tablehead{
	\colhead{ID} & \colhead{R.A.}      & \colhead{Decl.}   & \colhead{$z'$}  & \colhead{$V-I_{\mathrm{C}}$} & \colhead{$I_{\mathrm{C}}-z'$} & \colhead{Redshift} & \colhead{Reference$^a$} \\
	\colhead{}   & \colhead{(J2000.0)} & \colhead{(J2000.0)} & \colhead{(mag)} & \colhead{(mag)}  & \colhead{(mag)}   & \colhead{} & \colhead{}
}
\startdata
N146624(F36279-1750) & 12:36:27.74 & +62:17:47.8 & $26.05$ & $>$$2.32$ & $-0.11$ & $4.94$ & 1 \\
N119188(HDF 3-951.0) & 12:37:00.23 & +62:12:19.8 & $24.69$ & $1.82$  & $0.22$  & $5.34$ & 2 \\
N116886(HDF 4-625.0) & 12:36:44.65 & +62:11:50.7 & $25.24$ & $>$$2.87$ & $0.03$  & $4.58$ & 3 \\ 
N136110(ES1)         & 12:36:49.23 & +62:15:38.8 & $25.48$ & $>$$2.96$ & $-0.16$ & $5.19$ & 4 \\ 

N135557              & 12:36:55.48 & +62:15:32.9 & $25.64$ & $>$$2.56$ & $0.16$  & $5.19$ & 5 \\
N139345              & 12:38:06.84 & +62:16:20.4 & $24.89$ & $1.79$  & $-0.02$ & $4.91$ & 5 \\

N104268(A04-1)       & 12:38:11.32 & +62:09:19.4 & $24.02$ & $2.22$  & $0.11$  & $4.52$ & 6 \\
N144200(A04-2)       & 12:37:57.49 & +62:17:19.0 & $24.08$ & $2.40$  & $0.07$  & $4.70$ & 6 \\
N95819(A04-4)        & 12:37:05.68 & +62:07:43.3 & $24.50$ & $2.26$  & $0.15$  & $4.65$ & 6 \\
N139294(A04-5)       & 12:38:28.96 & +62:16:18.8 & $24.39$ & $1.96$  & $0.20$  & $4.67$ & 6 \\
N149472(A04-6)       & 12:38:25.52 & +62:18:19.7 & $24.87$ & $>$$3.02$ & $0.14$  & $4.86$ & 6 \\
N129178(A04-7)       & 12:38:04.36 & +62:14:19.7 & $24.29$ & $>$$3.28$ & $0.21$  & $5.18$ & 6 \\
N148198(A04-8)       & 12:38:16.63 & +62:18:05.3 & $24.50$ & $>$$3.19$ & $0.42$  & $4.62$ & 6 \\

N106944              & 12:37:58.13 & +62:09:51.3 & $24.55$ & $2.32$  & $0.02$  & $4.64$ & 7 \\
N127245              & 12:36:35.49 & +62:13:50.3 & $24.89$ & $1.96$  & $0.07$  & $4.42$ & 7 \\
N141368              & 12:36:14.21 & +62:16:43.3 & $25.32$ & $>$$2.21$ & $0.21$  & $5.15$ & 7 \\

\hline

S64032(CDFb-G05)     & 00:53:51.28 & +12:24:21.3 & $24.31$ & $2.70$  & $0.29$  & $4.49$ & 8 \\

S106426(A06-1)       & 00:52:21.34 & +12:32:35.3 & $24.07$ & $2.59$  & $0.23$  & $4.80$ & 6 \\
S104115(A06-2)       & 00:52:43.27 & +12:32:08.2 & $24.22$ & $1.75$  & $0.16$  & $4.27$ & 6 \\
S091813(A06-3)       & 00:52:39.88 & +12:29:44.1 & $25.03$ & $1.98$  & $0.06$  & $4.39$ & 6 \\
S093014(A06-4)       & 00:52:37.37 & +12:29:58.7 & $25.26$ & $>$$1.85$ & $0.16$  & $4.49$ & 6 \\

S101900(CDFa-GD7)    & 00:53:35.57 & +12:31:44.1 & $23.72$ & $1.67$  & $0.17$  & $4.61$ & 7 \\
S103759(CDFa-G1)     & 00:53:33.21 & +12:32:07.3 & $23.54$ & $2.88$  & $0.03$  & $4.82$ & 7 \\
\enddata
\tablenotetext{a}{1: \citet{daw01}, 2: \citet{spi98}, 3: \citet{fer01}, 4: \citet{daw02}, 5: \citet{bar08}, 6: \citet{and04,and07}, 7: this study, 8: \citet{ste99}.}
\end{deluxetable}
\begin{deluxetable}{lccccccc}
\tabletypesize{\scriptsize}
\tablecaption{Spectroscopically identified galaxies ($z>4.2$) out of the color selection by \citet{iwa07}\label{speclbg2}}
\tablewidth{0pt}
\tablecolumns{8}
\tablehead{
	\colhead{ID} & \colhead{R.A.}      & \colhead{Decl.}   & \colhead{$z'$}  & \colhead{$V-I_{\mathrm{C}}$} & \colhead{$I_{\mathrm{C}}-z'$} & \colhead{Redshift} & \colhead{Reference$^a$} \\
	\colhead{}   & \colhead{(J2000.0)} & \colhead{(J2000.0)} & \colhead{(mag)} & \colhead{(mag)}  & \colhead{(mag)}   & \colhead{} & \colhead{}
}
\startdata
F36219-1516               & 12:36:21.88 & +62:15:17.0 & $25.39$ & $0.67$ &  $0.09$ & $4.89$ & 1 \\
F36376-1453               & 12:36:37.62 & +62:14:53.8 & $21.97$ & $3.25$ &  $0.78$ & $4.89$ & 1 \\
HDF 4-439.0               & 12:36:43.84 & +62:12:41.7 & $25.29$ & $1.48$ & $-0.19$ & $4.54$ & 2 \\
GOODS J123721.03+621502.1 & 12:37:21.00 & +62:15:02.1 & $23.32$ & $2.36$ &  $0.56$ & $4.76$ & 3 \\
\enddata
\tablenotetext{a}{1: \citet{daw01}, 2: \citet{ster99}, 3: \citet{cow04}}
\end{deluxetable}

\begin{figure}[!t]
\begin{center}
\epsscale{0.7}
\plotone{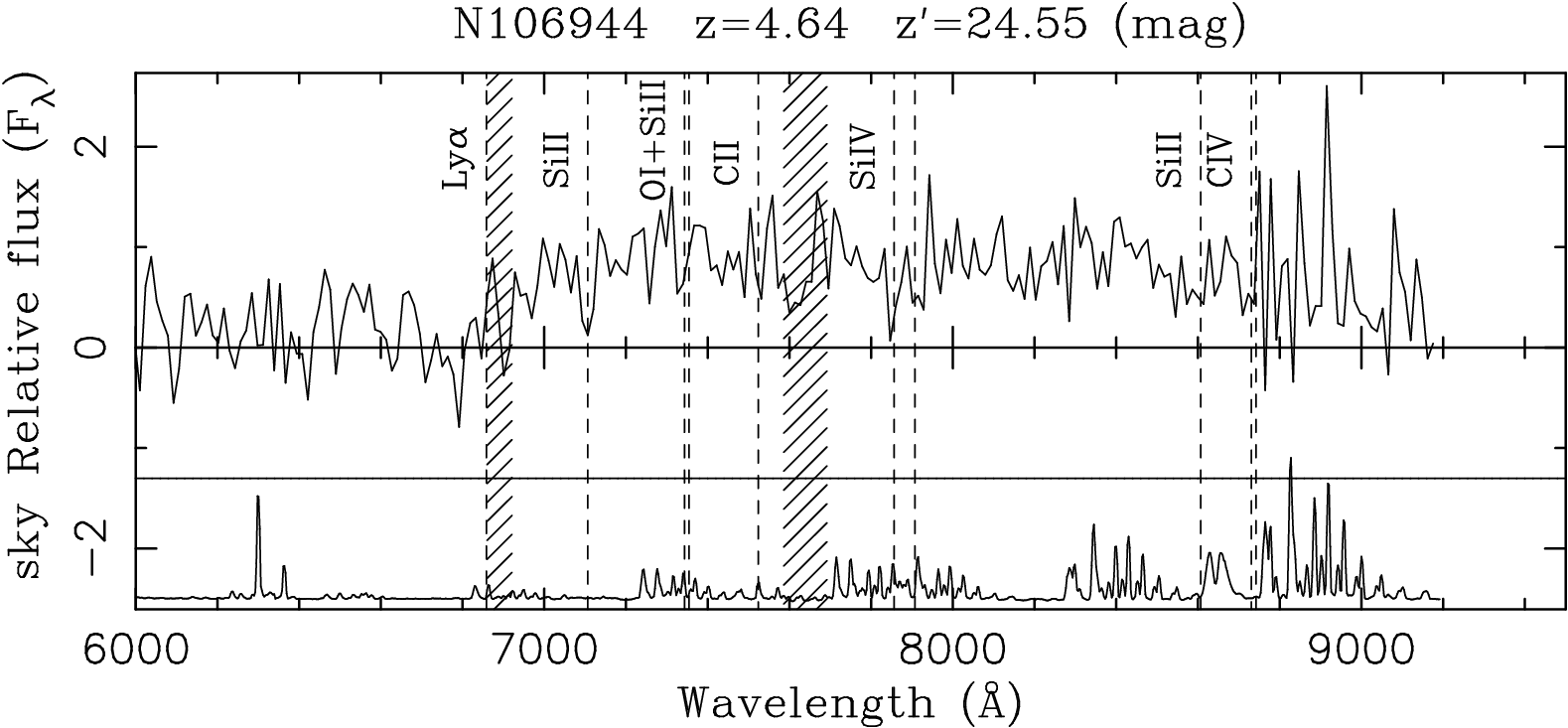}	

\plotone{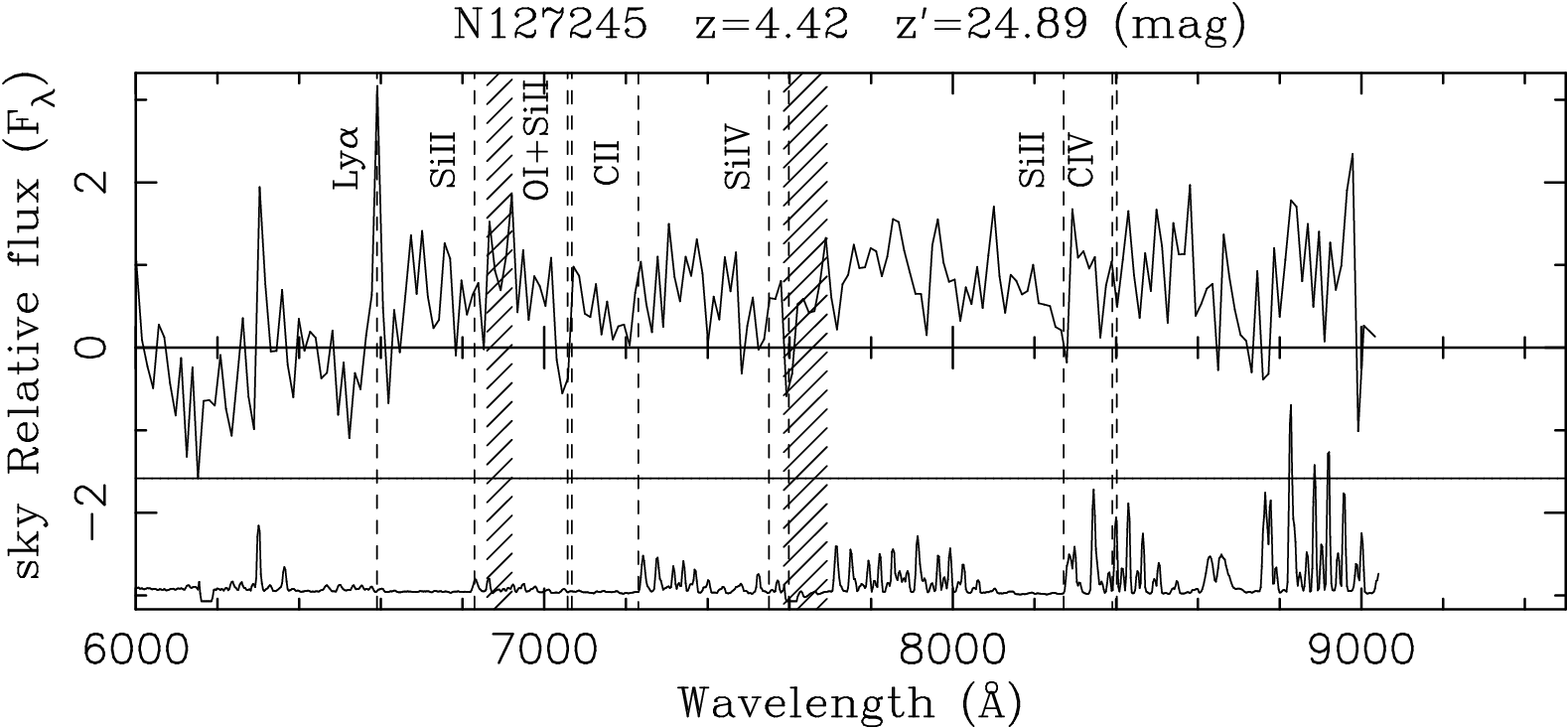}

\plotone{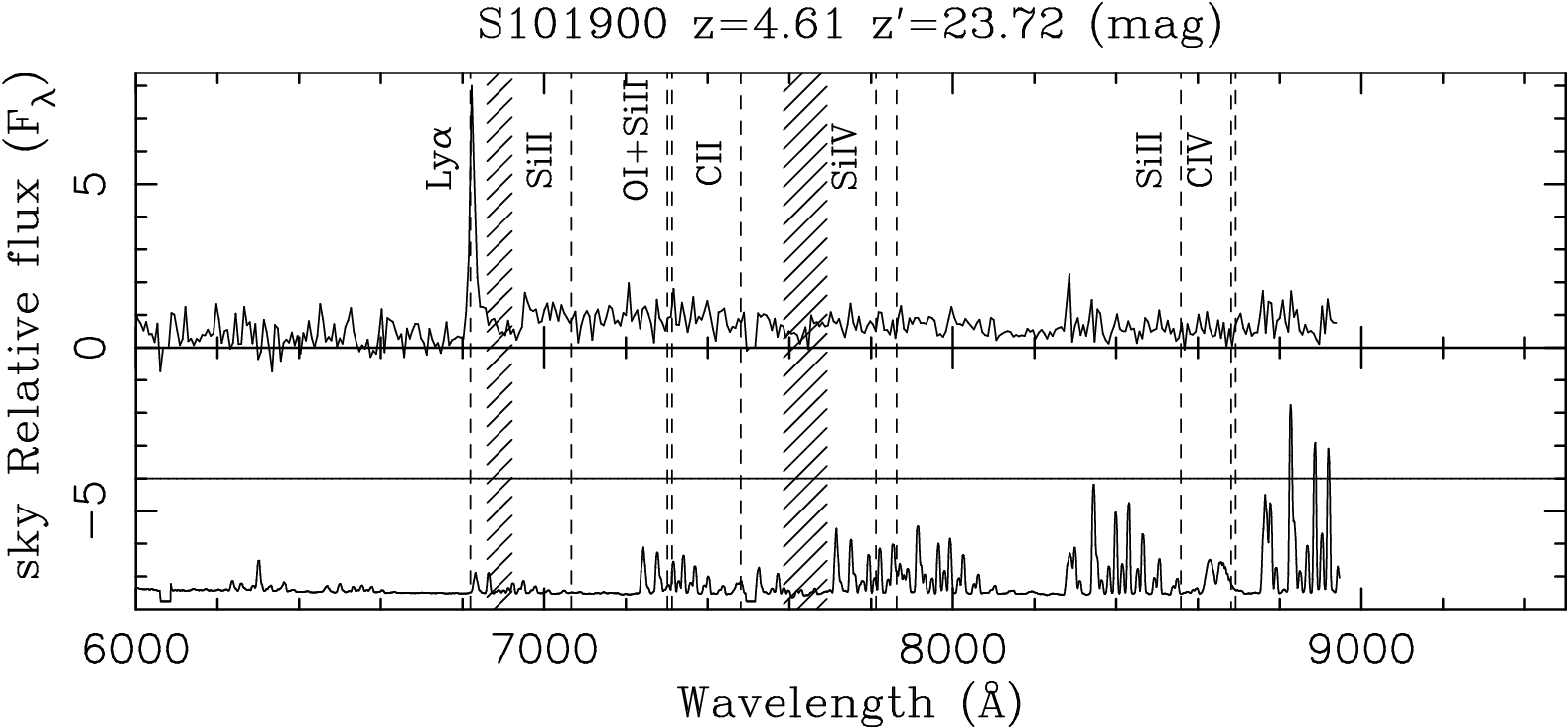}

\end{center}
\caption{
	Spectrum of observed $z \sim 5$ LBGs identified. {\it Top}: N106944. {\it Middle}: N127245. {\it Bottom}: S101900.
	The positions of redshifted Ly$\alpha$ and interstellar absorption lines are shown with vertical dashed lines.
	The sky spectrum is shown in the lower part  of each panel.
	The atmospheric A-band and B-band absorptions are shown with vertical hatched regions.
}
\label{spectrum}
\end{figure}

\clearpage

\begin{figure}[!t]
\begin{center}
\setcounter{figure}{0}
\epsscale{0.7}
\plotone{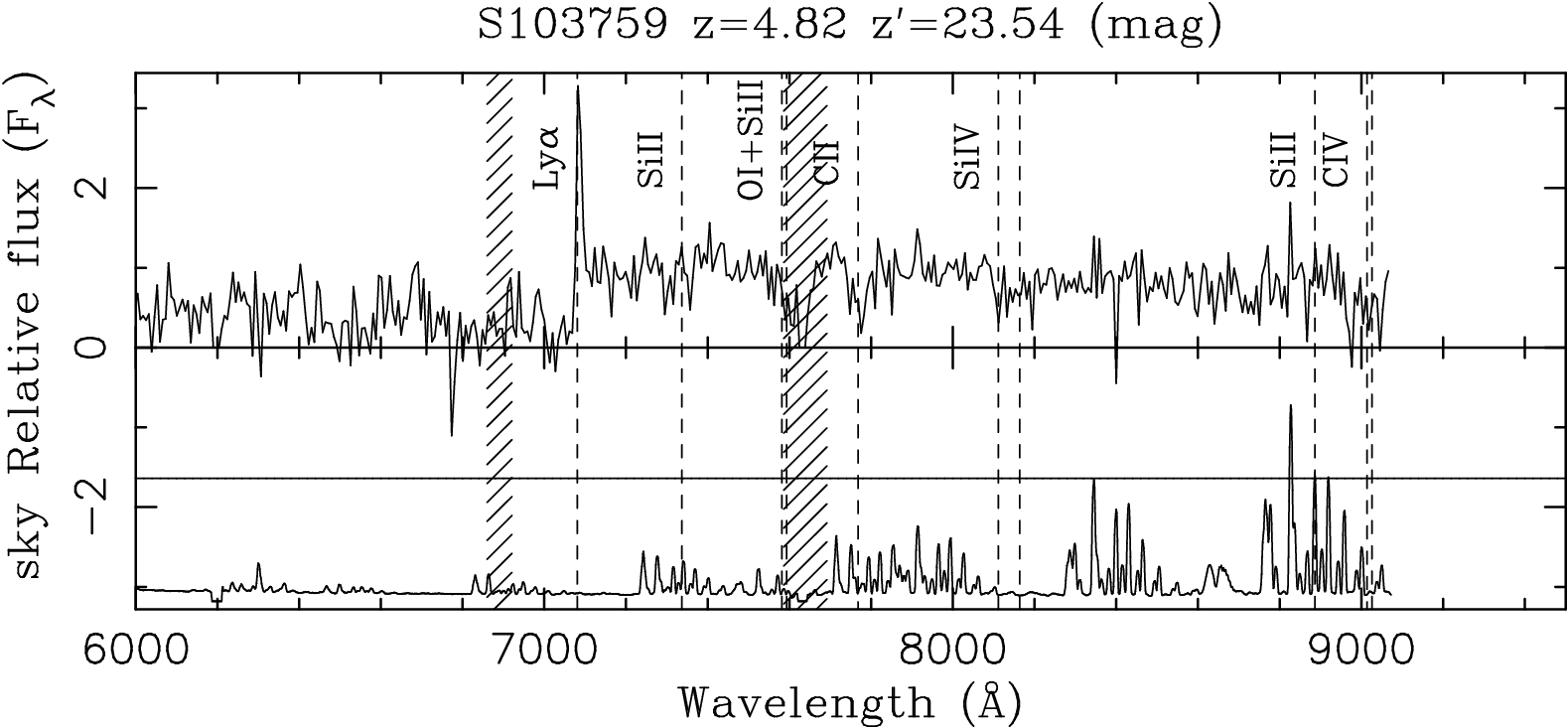}

\plotone{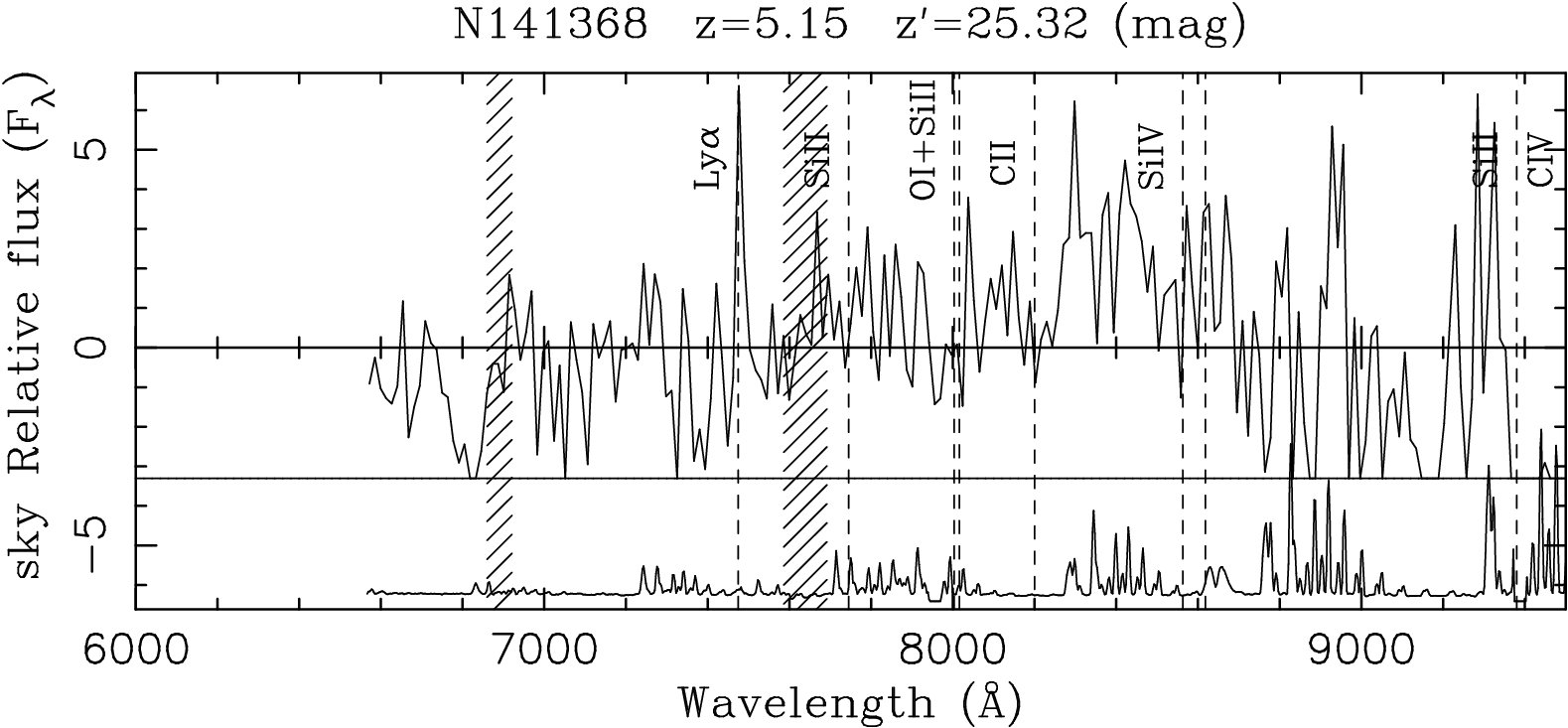}
\caption{
{\it Continued.} {\it Top}: S103759. {\it Bottom}: N141368.}
\end{center}
\end{figure}

\clearpage

\end{document}